\documentclass[twocolumn,aps]{revtex4}

\usepackage{graphicx}

\newcommand{\bphi}{{\mbox{\boldmath $\varphi$}}}

\begin{document}

\title{Asymmetry and decoherence in a double-layer persistent-current qubit}

\author{Guido Burkard}
\altaffiliation{Present address: 
Department of Physics and Astronomy,
University of Basel,
Klingelbergstrasse 82,
CH-4056 Basel, Switzerland}
\author{David P. DiVincenzo}
\affiliation{IBM T.\ J.\ Watson Research Center,
         P.\ O.\ Box 218,
         Yorktown Heights, NY 10598, USA}
\author{P. Bertet}
\author{I. Chiorescu}
\altaffiliation{Present address: 
National High Magnetic Field Laboratory,
Florida State University,
1800 East Paul Dirac Drive,
Tallahassee, FL 32310, USA}
\author{J. E. Mooij}
\affiliation{Quantum Transport Group, Kavli Institute of
Nanoscience, Delft University of Technology, Lorentzweg 1,
2628CJ, Delft, The Netherlands}

\begin{abstract}
Superconducting circuits fabricated using the widely used shadow evaporation technique can contain unintended junctions which change their quantum dynamics.  We discuss a superconducting flux qubit design that exploits the symmetries of a circuit to protect the qubit from unwanted coupling to the noisy environment, in which the unintended junctions can spoil the quantum coherence.  We present a theoretical model based on a recently developed circuit theory for superconducting qubits and calculate relaxation and decoherence times that can be compared with existing experiments.  Furthermore, the coupling of the qubit to a circuit resonance (plasmon mode) is explained in terms of the asymmetry of the circuit.  Finally, possibilities for prolonging the relaxation and decoherence times of the studied superconducting qubit are proposed on the basis of the obtained results.
\end{abstract}

\maketitle

\section{Introduction}
Superconducting (SC) circuits in the regime where the Josephson energy $E_J$
dominates the charging energy $E_C$ represent one of the currently 
studied candidates for a solid-state qubit \cite{MSS}.
Several experiments have demonstrated the quantum coherent behavior
of a SC flux qubit \cite{Mooij,Orlando,vanderWal}, 
and recently, coherent free-induction
decay (Ramsey fringe) oscillations have been observed \cite{CNHM}.
The coherence time $T_2$ extracted from these data was reported to 
be around $20\,{\rm ns}$, somewhat shorter than expected from 
theoretical estimates \cite{Devoret,Tian99,Tian02,WWHM}.
In more recent experiments \cite{Delft-new}, it was found that
the decoherence time $T_2$ can be increased up to approximately
$120\,{\rm ns}$ by applying a large dc bias current (about $80\%$
of the SQUID junctions' critical current).

A number of decoherence mechanisms can be important, being
both intrinsic to the Josephson junctions, e.g., oxide barrier 
defects \cite{Martinis} or
vortex motion, and external, e.g., current 
fluctuations from the external control circuits, e.g., 
current sources \cite{Devoret,Tian99,Tian02,WWHM,BKD}.
Here, we concentrate on the latter effect, i.e., current fluctuations,
and use a recently developed circuit theory \cite{BKD} to analyze
the circuit studied in the experiment \cite{CNHM}. 

\begin{figure}
\centerline{\includegraphics[width=7cm]{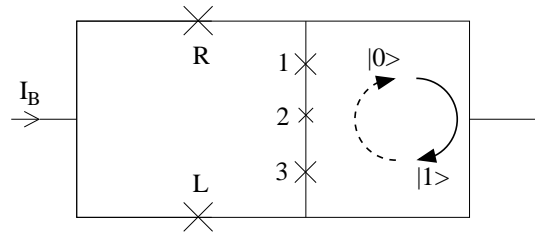}}
\caption{\label{qubit-i}
Schematic of the circuit.  Crosses denote Josephson junctions.
The outer loop with two junctions $L$ and $R$ forms a dc SQUID that
is used to read out the qubit.  The state of the qubit is determined 
by the orientation of the circulating current in the small loop,
comprising the junctions $1$, $2$, and $3$, one of which has a
slightly smaller critical current than the others.
A bias current $I_B$ can be applied as indicated for read-out.}
\end{figure}
The SC circuit studied in Ref.~\cite{CNHM} (see Fig.~\ref{qubit-i})
is designed to be 
immune to current fluctuations from the current bias line
due to its symmetry properties;  at zero dc bias, $I_B=0$,
and independent of the applied magnetic field,
a small fluctuating current $\delta I_B(t)$ caused by the
finite impedance of the external control circuit (the current 
source) is divided equally into the two arms of the 
SQUID loop and no net current flows through the
three-junction qubit line.  
\begin{figure}[b]
\centerline{\includegraphics[width=8cm]{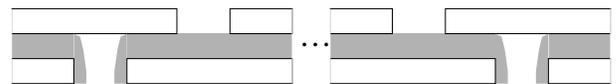}}
\caption{\label{shadow}
Schematics of Josephson junctions produced by the shadow 
evaporation technique, connecting the upper with the lower
aluminum layer.  Shaded regions represent the aluminum oxide.}
\end{figure}
Thus, in the ideal circuit, Fig.~\ref{qubit-i}, the qubit is 
protected from decoherence due to current fluctuations
in the bias current line.  This result also follows from
a systematic analysis of the circuit \cite{BKD}.
However, asymmetries in the SQUID loop may spoil the 
protection of the qubit from decoherence.  The breaking
of the SQUID's symmetry has other very interesting consequences,
notably the possibility to couple the qubit to an external
harmonic oscillator (plasmon mode) and thus to entangle the 
qubit with another degree of freedom \cite{Chiorescu-coupling}.
For an inductively coupled SQUID \cite{Mooij,Orlando,vanderWal}, 
a small geometrical
asymmetry, i.e., a small imbalance of self-inductances in a
SQUID loop combined with the same imbalance for the mutual
inductance to the qubit, is not sufficient to cause decoherence
at zero bias current \cite{BKD}.   A junction asymmetry,
i.e., a difference in critical currents in the SQUID
junctions $L$ and $R$, would in principle suffice to cause decoherence at
zero bias current.  However, in practice, the SQUID junctions
are typically large in area and thus their critical currents 
are rather well-controlled (in the system studied in 
Ref.~\cite{Delft-new}, the junction asymmetry is
$<5\%$), therefore the latter effect turns
out to be too small to explain the experimental findings.

An important insight in the understanding of decoherence in the 
circuit design proposed in \cite{CNHM} is that it
contains another asymmetry, caused by its double layer 
structure.  The double layer structure is an artifact of 
the fabrication method used to produce SC circuits
with aluminum/aluminum oxide Josephson junctions, the
so-called shadow evaporation technique.  Junctions produced
with this technique will always connect 
the top layer with the bottom layer, see Fig.~\ref{shadow}.   
\begin{figure}[b]
\centerline{\includegraphics[width=7.6cm]{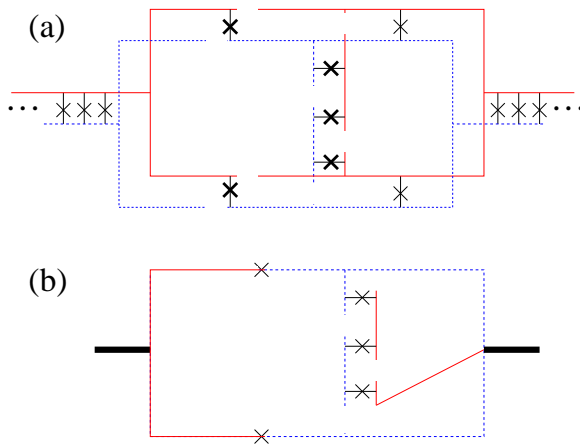}}
\caption{\label{qubit}
(a) Double layer structure.  Dashed blue lines represent the lower,
solid red lines the upper SC layer, and crosses indicate Josephson junctions.
The thick crosses are the intended junctions, while the thin
crosses are the unintended distributed junctions due to the double-layer
structure.
(b) Simplest circuit model of the double layer structure.
The symmetry between the upper and lower arms of the SQUID
has been broken by the qubit line comprising three junctions.
Thick black lines denote pieces of the SC in which the upper and lower
layer are connected by large area junctions.}
\end{figure}
Thus, while circuits like 
Fig.~\ref{qubit-i} can be produced with this technique,
strictly speaking, loops will always contain an
even number of junctions.
In order to analyze the implications of the double layer
structure for the circuit in Fig.~\ref{qubit-i}, we draw
the circuit again, see Fig.~\ref{qubit}(a), but this time with separate upper and 
lower layers.  Every piece of the upper layer will be connected 
with the underlying piece of the lower layer via an ``unintentional'' Josephson junction.
\begin{figure}
\centerline{\includegraphics[width=7cm]{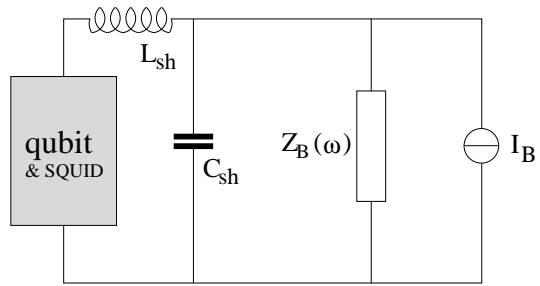}}
\caption{\label{system}
External circuit attached to the qubit (Fig.~\ref{qubit-i})
that allows the application of a bias current $I_B$ for qubit read-out.
The inductance $L_{\rm sh}$ and capacitance $C_{\rm sh}$
form the \textit{shell circuit}, and $Z(\omega)$ is the
total impedance of the current source ($I_B$).  The case
where a voltage source is used to generate a current
can be reduced to this using Norton's theorem.}
\end{figure}
However, these extra junctions typically have large areas 
and therefore large critical currents;  thus, their
Josephson energy can often be neglected.
Since we are only interested in the lowest-order effect of the
double layer structure, we neglect all unintentional junctions in this
sense;  therefore, we arrive at the circuit, Fig.~\ref{qubit}(b),
without extra junctions.  We notice however, that this
resulting circuit is distinct from the 'ideal' circuit
Fig.~\ref{qubit-i} that does not reflect the double-layer
structure.  In the real circuit,  Fig.~\ref{qubit}(b), the
symmetry between the two arms of the dc SQUID is 
broken, and thus it can be expected that
bias current fluctuations cause decoherence of the
qubit at zero dc bias current, $I_B=0$.  
This effect is particularly important in the circuit discussed in
\cite{CNHM,Delft-new} since the coupling between the qubit and the 
SQUID is dominated by the kinetic inductance of the shared line, 
and so is strongly asymmetric, 
rather than by the geometric mutual inductance \cite{vanderWal}
which is symmetric.
Our analysis below will show this quantitatively and will allow us 
to compare our theoretical predictions with 
the experimental data for the decoherence times as
a function of the bias current.  Furthermore,
we will theoretically explain the coupling of the qubit to 
a plasma mode in the read-out circuit (SQUID plus external circut),
see Fig.~\ref{system},
at $I_B=0$;  this coupling is absent for a symmetric circuit.

This article is organized as follows.  In Sec.~\ref{Hamiltonian},
we derive the Hamiltonian of the qubit, taking into account
its double-layer structure.  We use this Hamiltonian to calculate 
the relaxation and decoherence times as a function of the applied bias current
(Sec.~\ref{Decoherence}) and to derive an effective Hamiltonian for the
coupling of the qubit to a plasmon mode in the read-out circuit (Sec.~\ref{Plasmon}).
Finally, Sec.~\ref{Discussion} contains a short discussion of our result
and possible lessons for future SC qubit designs.

\section{Hamiltonian}
\label{Hamiltonian}

\begin{figure}[t]
\centerline{\includegraphics[width=7cm]{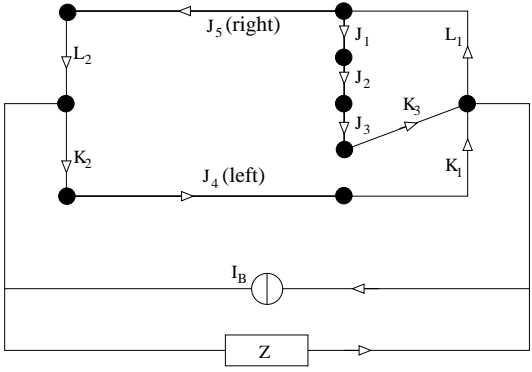}}
\caption{\label{graph}
The network graph of the circuit, Figs.~\ref{qubit}b and \ref{system}.
Dots indicate the nodes, lines the branches of the
graph;  an arrow indicates the  orientation of a branch. 
Thick lines labeled $J_i$ denote an RSJ element, i.e., a
Josephson junction shunted by a capacitor and a resistor.
Lines labeled $L_i$ and $K_i$ denote inductances,
$Z_{\rm ext}$ the external impedance, including the shell circuit
of Fig.~\ref{system}, and $I_B$ is the current source.}
\end{figure}

\begin{figure}[t]
\centerline{\includegraphics[width=6cm]{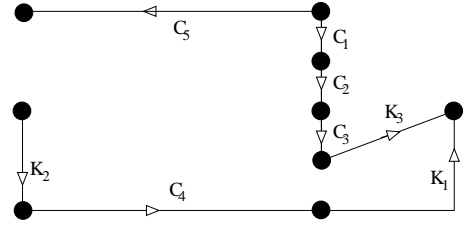}}
\caption{\label{tree}
A tree of the circuit graph, Fig.~\ref{graph}.
A tree is a subgraph connecting all nodes, containing
no loops.  Here, the tree was chosen to contain all
capacitors $C_i$ (from the RSJ elements) and as few
inductors $K_i$ as possible (see Ref.~\cite{BKD}).}
\end{figure}

In order to model the decoherence of the qubit, we need to find its 
Hamiltonian and its coupling to the environment.
The Hamiltonian of the circuit Fig.~\ref{qubit}(b) can be found using the
circuit theory developed in Ref.~\cite{BKD}.
To this end, we first draw the circuit graph (Fig.~\ref{graph}) and find a
tree of the circuit graph containing all capacitors and as few inductors
as possible (Fig.~\ref{tree}).
A tree of a graph is a subgraph containing all of its nodes but no loops.
By identifying the fundamental loops \cite{BKD} in the circuit graph
(Fig.~\ref{graph}) we obtain the loop submatrices
\begin{eqnarray}
  \label{F}
  {\bf F}_{CL} = \left(\begin{array}{r r}
   -1 & 1 \cr -1 & 1 \cr -1 & 1 \cr 0 & -1 \cr 0 & -1\end{array}\right),
  & & {\bf F}_{CZ} = -{\bf F}_{CB} = \left(\begin{array}{r} 0 \cr 0 \cr 0 \cr 1 \cr 0\end{array}\right),\\
  {\bf F}_{KL} = \left(\begin{array}{r r} 0 & -1 \cr 0 & -1 \cr -1 & 1 \end{array}\right),
  & & {\bf F}_{KZ} = -{\bf F}_{KB} = \left(\begin{array}{r} 1 \cr 1 \cr 0 \end{array}\right).
\end{eqnarray}
The chord ($L$) and tree ($K$) inductance matrices are taken to be
\begin{equation}
  \label{L}
  {\bf L} = \left(\begin{array}{c c}
      L/2 & M/4 \\
      M/4 & L'/2 
\end{array}\right), \quad \quad
  {\bf L}_K =\left(\begin{array}{c c c}
      L/2 & M/4  & M_i \\
      M/4 & L'/2 & 0\\
      M_i & 0    & L_i
\end{array}\right),
\end{equation}
where $L$, $L'$, and $L_i$ are, respectively, the self-inductances of the qubit loop
in the upper layer, the SQUID, and qubit loop in the lower layer,
and $M$ and $M_i$ are the mutual inductances between the qubit and the SQUID
and between the upper and lower layers in the qubit loop.
The tree-chord mutual inductance matrix is taken to be
\begin{equation}
  {\bf L}_{LK} = \left(\begin{array}{c c c}
      0   & M/4 & 0\\
      M/4 & 0   & 0
\end{array}\right).
\end{equation}
The Hamiltonian in terms of the SC phase differences 
$\bphi=(\varphi_1,\varphi_2,\varphi_3,\varphi_L,\varphi_R)$ across the Josephson 
junctions and their conjugate variables, the capacitor charges ${\bf Q}_C$,
is found to be \cite{BKD}
\begin{equation}
  {\cal H}_S  =   \frac{1}{2}{\bf Q}_C^T {\bf C}^{-1}{\bf Q}_C
                + \left(\frac{\Phi_0}{2\pi}\right)^2  U(\bphi),  \label{Hamiltonian-S}
\end{equation}
with the potential
\begin{eqnarray}
U(\bphi) &=& -\sum_i \frac{1}{L_{J;i}}\cos\varphi_i
             + \frac{1}{2L_Q}\left(\varphi_1+\varphi_2+\varphi_3-f\right)^2\nonumber\\
         &+& \frac{1}{2L_S}\left(\varphi_L+\varphi_R-f'\right)^2 \label{U}\\
         &+& \frac{1}{M_{QS}}\left(\varphi_1+\varphi_2+\varphi_3-f\right)
                                \left(\varphi_L+\varphi_R-f'\right)\nonumber \\
         &+& \frac{2\pi}{\Phi_0}I_B\left[m_Q\left(\varphi_1+\varphi_2+\varphi_3\right)
               + m_L\varphi_L+m_R\varphi_R \right], \nonumber
\end{eqnarray}
where the Josephson inductances are given by $L_{J;i} = \Phi_0/2\pi I_{c;i}$, 
and $I_{c;i}$ is the critical current of the $i$-th junction.
In Eq.~(\ref{U}), we have also introduced the effective self-inductances
of the qubit and SQUID and the effective qubit-SQUID mutual inductance,
\begin{eqnarray}
  \label{Hinductances}
  L_Q &=& L \frac{\kappa}{4(1+L'/L+2M/L)},\\
  L_S &=&  L \frac{\kappa}{2(1+2L_i/L)},\\ 
  M_{QS} &=& -L \frac{\kappa}{2(1+M/L+2M_i/L)},
\end{eqnarray}
and the coupling constants between the bias current
and the qubit and the left and right SQUID phases,
\begin{eqnarray}
  m_Q &=& \kappa^{-1} (1+L'/L+2M/L)(1-2M_i/L), \\
  m_L &=&  \frac{1}{2}-\frac{1}{2\delta}, \quad\quad
  m_R = -\frac{1}{2}-\frac{1}{2\delta},
\end{eqnarray}
with the definitions
\begin{eqnarray}
\kappa &=& 1+4L_i(L+L'+2M)/L^2+2(L'+M-2M_i)/L\nonumber\\
&&-(M+2M_i)^2/L^2,\label{kappa}\\
\delta &=& \kappa/(1+M/L+2M_i/L)(1-2M_i/L).\label{delta}
\end{eqnarray}
The sum $\varphi_1+\varphi_2+\varphi_3$ is the total phase difference across the qubit
line containing junctions $J_1$, $J_2$, and $J_3$, whereas
$\varphi_L+\varphi_R$ is the sum of the phase differences in the SQUID loop.
Furthermore, ${\bf C}={\rm diag}(C,C,C,C',C')$ is the capacitance matrix,
$C$ and $C'$ being the capacitances of the qubit and SQUID junctions, respectively.

The working point is given by the triple $(f,f',I_B)$, i.e., by the bias current $I_B$, 
and by the dimensionless external magnetic fluxes threading the qubit and SQUID loops, 
$f=2\pi\Phi_x/\Phi_0$ and $f'=2\pi\Phi_x'/\Phi_0$.
We will work in a region of parameter space where the potential $U(\bphi)$ has
a double-well shape, which will be used to encode the logical
qubit states $|0\rangle$ and $|1\rangle$.

The classical equations of motion, including dissipation, are
\begin{equation}
  \label{class-eqm}
  {\bf C} \ddot\bphi = -\frac{\partial U}{\partial \bphi} -\mu K*{\bf m}({\bf m}\cdot\bphi),
\end{equation}
where convolution is defined as $(f*g)(t) = \int_{-\infty}^t f(t-\tau)g(\tau)d\tau$.
The vector ${\bf m}$ is given by
\begin{equation}
  \label{m}
  {\bf m}=A(m_Q,m_Q,m_Q,m_L,m_R),
\end{equation}
and $A$ is chosen such that $|{\bf m}|=1$.
For the coupling constant $\mu$, we find
\begin{eqnarray}
  \mu  &=&  \kappa^{-2}L^{-4}
            \left(3\,{\left( {L} + {L'} + 2\,M \right) }^2\,{\left( {L} + M - 2\,{M_i} \right) }^2 \right.\nonumber\\ 
       & & \left. +\left( 2\,{L_i}\,\left( {L} + {L'} + 2\,M \right)  
           + {L}\,\left( {L'} - 2\,{M_i} \right)  \right.\right. \label{mu}\\ 
       & &  \left.\left. - M\,\left( M + 2\,{M_i} \right)  \right)^2
        +    \left( {{L}}^2 + 2\,{L_i}\,\left( {L} + {L'} + 2\,M \right)  \right.\right. \nonumber\\ 
      & &  \left. \left. + {L} \,\left( {L'} + 2\,M - 2\,{M_i} \right)  - 2\,{M_i}\,\left( M + 2\,{M_i} \right)\right)^2\right).\nonumber
\end{eqnarray}
The kernel $K$ in the dissipative term is determined by the total external
impedance;  in the frequency domain,
\begin{equation}
  \label{K}
  K(\omega) =  \frac{i\omega}{Z(\omega)},
\end{equation}
with the impedance
\begin{equation}
  \label{Z}
  Z(\omega) =  Z_{\rm ext}(\omega) + i\omega L_{\rm int},
\end{equation}
where we have defined the internal inductance,
\begin{eqnarray}
  L_{\rm int} &=& \frac{1}{4\kappa L^2} \left(
  4 L_i ( L + L' ) ( L + L' + 2 M ) + 2 L^2 L' \right.\nonumber\\
  & & - L M^2   -  4 L' M M_i - 8 L' M_i^2 - 8 M M_i^2\\
  & & \left. + L ( 2 L'^2 + 2 L' M - M^2 + 4 M M_i - 8 M_i^2)\right),\nonumber
\end{eqnarray}
and where
\begin{equation}
  \label{Zext}
  Z_{\rm ext} = \left(\frac{1}{Z_B(\omega)}+i\omega C_{\rm sh}\right)^{-1}
                +i\omega L_{\rm sh}
\end{equation}
is the impedance of the external circuit attached to the
qubit, including the shell circuit, see Figs.~\ref{system},\ref{graph}.
For the parameter regime we are interested in, $L_{\rm int}\approx 20\,{\rm pH}$,
$\omega\lesssim 10\,{\rm GHz}$, and $Z\gtrsim 50\,\Omega$ 
therefore $\omega L_{\rm int}\ll |Z_{\rm ext}|$,
and we can use $Z(\omega)\approx Z_{\rm ext}(\omega)$.

\begin{figure}[t]
\centerline{\includegraphics[width=8.5cm]{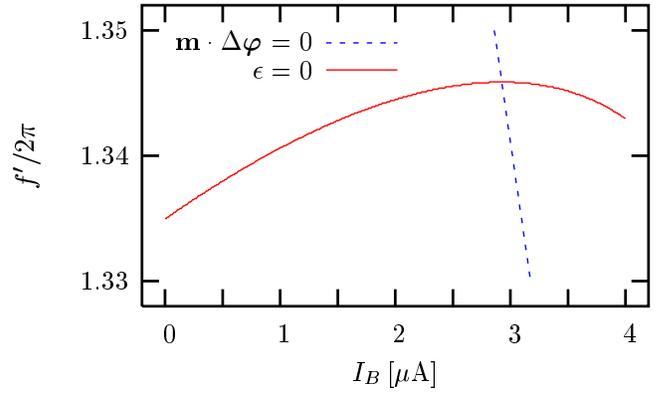}}
\caption{\label{decoupling}
Decoupling (red solid) and symmetric (blue dashed) curves in the $(I_B,f')$ plane,
where $I_B$ is the applied bias current and $f'=2\pi \Phi_x'/\Phi_0$
is the dimensionless externally applied magnetic flux threading the SQUID loop.
Both curves are obtained from the numerical minimization of the potential
Eq.~(\ref{U}).  The decoupling line is determined using the condition 
${\bf m}\cdot\Delta\bphi=0$, whereas the symmetric line follows from the
condiction $\epsilon=0$.}
\end{figure}

We numerically find the double-well minima $\bphi_0$ and $\bphi_1$
for a range of bias currents between $0$ and $4\,\mu{\rm A}$ and
external flux $f'/2\pi$ between $1.33$ and $1.35$ and a qubit flux
around $f/2\pi\simeq 0.5$ (the ratio $f/f'=0.395$ is fixed by the 
areas of the SQUID and qubit loops in the circuit).  
The states
localized at $\bphi_0$ and $\bphi_1$ are encoding the logical $|0\rangle$
and $|1\rangle$ states of the qubit. This allows us to find the set of parameters
for which the double well is symmetric, $\epsilon \equiv
U(\bphi_0)-U(\bphi_1)=0$.  The curve $f^*(I_B)$ on which 
the double well is symmetric is plotted in Fig.~\ref{decoupling}.
Qualitatively, $f^*(I_B)$ agrees well with the experimentally measured
symmetry line \cite{Delft-new}, but it underestimates the magnitude
of the variation in flux $f'$ as a function of $I_B$.
The value of $I_B$ where the symmetric and the decoupling lines intersect
coincides with the maximum of the symmetric line, as can be understood
from the following argument.  Taking the total derivative with respect 
to $I_B$ of the relation 
$\epsilon = U(\bphi_0;f^*(I_B),I_B)-U(\bphi_1;f^*(I_B),I_B)=0$ 
on the symmetric line, and using that $\bphi_{0,1}$ are extremal
points of $U$, we obtain ${\bf n}\cdot\Delta\bphi \,\partial f^*/\partial I_B
+(2\pi/\Phi_0){\bf m}\cdot\Delta\bphi=0$ for some constant vector ${\bf n}$.
Therefore, ${\bf m}\cdot\Delta\bphi=0$ (decoupling line) and 
${\bf n}\cdot\Delta\bphi\neq 0$ implies $\partial f^*/\partial I_B =0$.

For the numerical calculations throughout this paper, we use
the estimated experimental parameters from \cite{Delft-new,Chiorescu-coupling},
$L=25\,{\rm pH}$, $L'=45\,{\rm pH}$, $M=7.5\,{\rm pH}$, $L_i=10\,{\rm pH}$, $M_i=4\,{\rm pH}$, 
$I_{c;L}=I_{c;R}=4.2\,\mu{\rm A}$, and $I_{c;1}=I_{c;2}/\alpha=I_{c;3}=0.5\,\mu{\rm A}$ 
with $\alpha\simeq 0.8$.

\section{Decoherence}
\label{Decoherence}

The dissipative quantum dynamics of the qubit will be described using a Caldeira-Leggett
model \cite{CaldeiraLeggett} which is consistent with the classical dissipative equation
of motion, Eq.~(\ref{class-eqm}).  
We then quantize the combined system and bath Hamiltonian and use the master equation for
the superconducting phases $\bphi$ of the qubit and SQUID in the Born-Markov approximation 
to obtain the relaxation and decoherence times of the qubit.

\subsection{Relaxation time $T_1$}
\label{relaxation}

\begin{figure}
\centerline{\includegraphics[width=8.5cm]{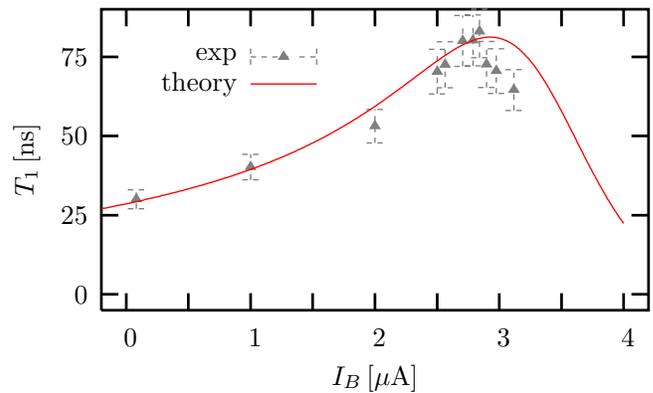}}
\caption{\label{times1}
Theoretical relaxation time $T_1$ (solid line)
as a function of the applied bias current $I_B$,
along the symmetric line (Fig.~\ref{decoupling}).
The value of $I_B$ where $T_1$ diverges coincides with the 
intersection of the symmetric line with the decoupling line 
in Fig.~\ref{decoupling};  the divergence is removed in the theory curve
by including a residual impedance of $R_{\rm res}=3.5\,{\rm M}\Omega$.
The experimentally obtained data for sample A in Ref.~\onlinecite{Delft-new}
are shown as triangle symbols with error bars.}
\end{figure}

\begin{figure}[b]
\centerline{\includegraphics[width=8.5cm]{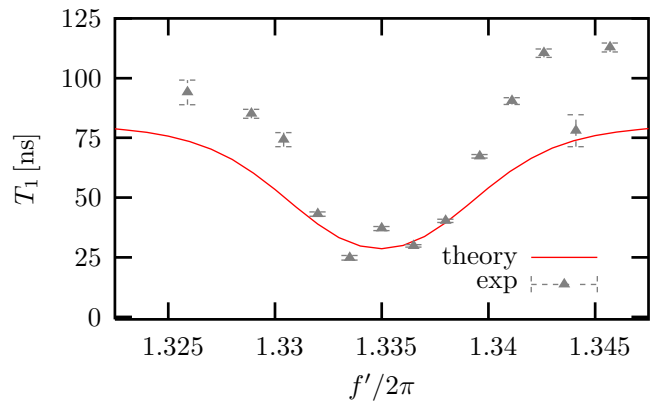}}
\caption{\label{times10}
Theoretical relaxation time $T_1$ (solid line)
as a function of the applied magnetic flux $f'=\Phi_x'/\Phi_0$ 
at zero bias current, $I_B=0$, around the symmetric point, $\epsilon=0$.
Experimentally obtained data for sample A in Ref.~\onlinecite{Delft-new}
are shown as triangle symbols with error bars.
The theory curve from the semiclassical $T_1$ formula, Eq.~(\ref{T1}),
is expected to be valid in the range $|\epsilon| \lesssim \Delta$,
which corresponds roughly to $1.33\lesssim f'/2\pi \lesssim 1.34$.
Experimental points outside the plotted range of $f'$ where the theory curve
is not expected to be valid, are not shown.}
\end{figure}

The relaxation time of the qubit in the semiclassical
approximation \cite{fn1} is given by
\begin{equation}
  T_1^{-1}     =  \frac{\Delta^2}{E^2}\left(\frac{\Phi_0}{2\pi}\right)^2
                   |{\bf m}\cdot\Delta\bphi|^2 {\rm Re}\frac{E}{Z(E)} 
                   \coth\left(\frac{E}{2k_B T}\right),  \label{T1}
\end{equation}
where $\Delta\bphi \equiv \bphi_0-\bphi_1$ is the vector joining the two minima in
configuration space and
\begin{equation}
  \label{splitting}
  E=\sqrt{\Delta^2 + \epsilon^2}
\end{equation}
is the energy splitting between the two (lowest) eigenstates of the double well
and $\Delta$ is the tunnel coupling between the two minima.
We will evaluate $T_1$ on the symmetric line where $\epsilon=0$, and therefore,
$E=\Delta$.
At the points in parameter space $(I_B,f')$ where ${\bf m}\cdot\Delta\bphi$
vanishes, the system will be decoupled from the environment (in lowest order perturbation theory),
and thus $T_1\rightarrow\infty$.  From our numerical determination of 
$\bphi_0$ and $\bphi_1$, the decoupling flux $f'$, at which 
${\bf m}\cdot\Delta\bphi=0$, is obtained as a function of $I_B$ (Fig.~\ref{decoupling}).
From this analysis, we can infer the parameters $(I_B,f')$
at which $T_1$ will be maximal and the relaxation time away from the divergence.  
In practice, the divergence will be cut off by other effects which lie beyond the 
scope of this theory.  However, we can fit the peak value
of $T_1$ from recent experiments \cite{Delft-new} with a residual 
impedance of $R_{\rm res}\simeq 3.5\,{\rm M}\Omega$ which lies in a
different part of the circuit than $Z$ (Fig.\ref{graph}).
We do not need to further specify the position of $R_{\rm res}$ in the circuit;
we only make use of the fact that it gives rise to an additional contribution to the relaxation
rate of the form Eq.~(\ref{T1}) but with a vector ${\bf m}_{\rm res}\neq {\bf m}$, 
with ${\bf m}_{\rm res}\cdot\Delta\bphi\neq 0$ on the decoupling line.
Without loss of generality, we can adjust $R_{\rm res}$ such that
${\bf m}_{\rm res}\cdot\Delta\bphi = 1$.
Such a residual coupling may, e.g., originate from the subgap resistances of the 
junctions.  The relaxation time $T_1$ obtained
from Eq.~(\ref{T1}) as a function of $I_B$ along the symmetric 
line $\epsilon=0$ (Fig.~\ref{decoupling}) with a cut-off of the divergence by $R_{\rm res}$ is plotted in Fig.~\ref{times1}, along with the experimental data from sample A in \cite{Delft-new}.  In Fig.~\ref{times10}, we also plot $T_1$ (theory and experiment) as a function of the applied magnetic flux around the symmetric point at zero bias current.
For the plots of $T_1$ in Figs.~\ref{times1} and \ref{times10},
we have used the experimental parameters $\Delta/h=5.9\,{\rm GHz}$, 
$Z(E)\simeq Z_{\rm ext}(E)=60\,\Omega$, and $T=100\,{\rm mK}$.

\subsection{Decoherence time $T_2$}
\label{dephasing}

\begin{figure}[t]
\centerline{\includegraphics[width=8.5cm]{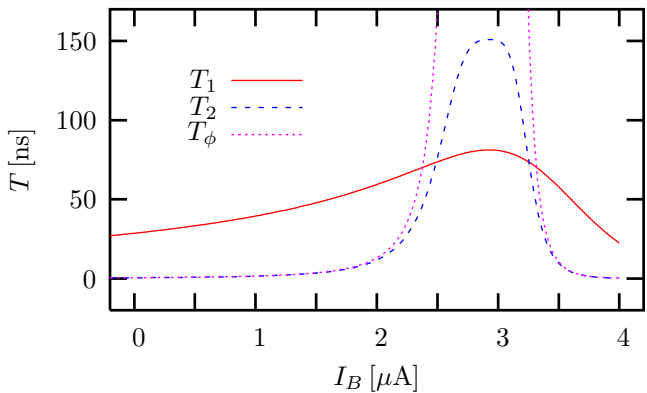}}
\caption{\label{times}
Theoretical relaxation, pure dephasing, and decoherence times $T_1$, $T_\phi$, and $T_2$
as a function of applied bias current $I_B$, along the symmetric line (Fig.~\ref{decoupling}).
As in Fig.~\ref{times1}, we have included decoherence from
a residual impedance of $R_{\rm res}=3.5\,{\rm M}\Omega$.}
\end{figure}
The decoherence time $T_2$ is related to the relaxation time $T_1$ via
\begin{equation}
  \label{T2}
  \frac{1}{T_2}  =   \frac{1}{T_\phi}  + \frac{1}{2 T_1},
\end{equation}
where $T_\phi$ denotes the (pure) dephasing time.
On the symmetric line $f'=f^*(I_B)$ (see Fig.~\ref{decoupling}), 
the contribution to the dephasing rate $T_{\phi}^{-1}$ 
of order $R_Q/Z$ vanishes, where 
$R_Q=e^2/h\approx 25.8\,{\rm k}\Omega$ denotes the quantum of resistance.
However, there is a second-order contribution $\propto (R_Q/Z)^2$, 
which we can estimate as follows.
The asymmetry $\epsilon = U(\bphi _0)-U(\bphi _1)$ 
of the double well as a function of the bias current $I_B$ at fixed
external flux $f'$ can be written in terms of a Taylor series around $I_B^*$,
\begin{equation}
  \label{Tphi-a1}
  \epsilon(I_B) = \epsilon_0 + \epsilon_1 \delta I_B + \epsilon_2 \delta I_B^2 + O(\delta I_B)^3,
\end{equation}
where $\delta I_B(t)=I_B(t)-I_B^*$ is the variation away from the dc bias current $I_B^*$.
The coefficients $\epsilon_i(I_B)$ can be obtained numerically from the minimization of the 
potential $U$, Eq.~(\ref{U}).  The approximate two-level Hamiltonian 
$\frac{\Delta}{2}\sigma_X + \frac{\epsilon}{2}\sigma_Z$
in its eigenbasis is then, up to $O(\delta I_B^3)$,
\begin{eqnarray}
  \label{twolevel}
  H  &=& \frac{1}{2}\sqrt{\Delta^2+\epsilon^2}\sigma_z
     =   \frac{\Delta}{2}\sigma_z  +  \frac{\epsilon^2}{4 \Delta} \sigma_z\\
     &=& \frac{\tilde\Delta}{2}\sigma_z  +  \frac{\epsilon_0 \epsilon_1}{2\Delta}\sigma_z\delta I_B
        + \left(\frac{\epsilon_1^2}{4\Delta}+  \frac{\epsilon_0 \epsilon_2}{2\Delta}\right)\sigma_z \delta I_B^2,
\quad %\quad
\end{eqnarray}
where $\tilde\Delta = \Delta + \epsilon_0^2/2\Delta$.
On the symmetric line, $\epsilon_0=0$, the term linear in $\delta I_B$
vanishes.  However, there is a non-vanishing second-order term  $\propto\epsilon_1^2$
that contributes to dephasing on the symmetric line.
Without making use of the correlators for $\delta I_B^2$,
we know that the pure dephasing rate $T_\phi^{-1}$ will be proportional
to $\epsilon_1(I_B)^4$ which allows us to predict the dependence of $T_\phi$ on $I_B$.
A discussion of the second-order dephasing within the
spin-boson model can be found in \cite{MakhlinShnirman}.  
However, in order to explain the order of magnitude of the experimental result \cite{Delft-new}
for $T_\phi$ correctly, the strong coupling to the plasma mode may also play an important 
role \cite{Delft-new,Bertet-unpublished}.
The result presented here cannot be used to predict the absolute magnitude of
$T_\phi$, but we can obtain an estimate for the dependence of $T_\phi$ on the
bias current $I_B$ via $\epsilon_1(I_B)=d\epsilon/dI_B$ obtained numerically
from our circuit theory, via
\begin{equation}
  T_\phi^{-1}(I_B) \approx T_\phi^{-1}(0)\left(\frac{\epsilon_1(I_B)}{\epsilon_1(0)}\right)^4,
      \label{Tphi}
\end{equation}
where for dimensional reasons we can write the proportionality constant in terms
of a zero-frequency resistance $R_0$ and an energy $\bar\omega$
(note, however, that this corresponds to one free parameter in the theory),
$T_\phi^{-1}(0)/\epsilon_1(0) \approx 2 \bar\omega^3/R_0^2\Delta^2$.
For the plots of $T_\phi$ and $T_2$ in Fig.~\ref{times},
we have used the resistance $R_0=1450\,\Omega$
and have chosen $\bar\omega/2\pi\approx 1\,{\rm THz}$ to
approximately fit the width of the $T_2$ curve.
The relaxation, dephasing, and decoherence times $T_1$, $T_\phi$, and $T_2$
are plotted as a function of the bias current $I_B$ in Fig.~\ref{times1}
and Fig.~\ref{times}.

The calculated relaxation and decoherence times $T_1$ and $T_2$ agree well with the experimental
data \cite{Delft-new} in their most important feature, the peak at
$I_B\approx 2.8\,\mu{\rm A}$.  This theoretical result does not involve fitting with any free 
parameters, since it follows exclusively from the independently known values for the circuit
inductances and critical currents. Moreover, we obtain good quantitative agreement between
theory and experiment for $T_1$ away from the divergence.
The shape of the $T_1$ and $T_2$ curves can be understood qualitatively from the theory.

\section{Coupling to the plasmon mode}
\label{Plasmon}

In addition to decoherence, the coupling to the external
circuit (Fig.~\ref{system})
can also lead to resonances in the microwave spectrum of the system
that originate from the coupling between the qubit to a LC resonator formed by the
SQUID, the inductance $L_{\rm sh}$ and capacitance $C_{\rm sh}$ of the ``shell'' 
circuit (plasmon mode).
We have studied this coupling quantitatively in the framework of the circuit 
theory \cite{BKD}, by replacing the circuit elements $I_B$ and $Z$ in the 
circuit graph by the elements $L_{\rm sh}$ and $C_{\rm sh}$ in 
series, obtaining the graph matrices
\begin{eqnarray}
  \label{F6}
  {\bf F}_{CL} = \left(\begin{array}{r r r} -1 & 1 & 0 \cr -1 & 1 & 0 \cr -1 & 1 & 0\cr 0 & -1 & 1\cr 0 & -1 & 0\cr 0 & 0 & -1 \end{array}\right)\!\!,\,
  {\bf F}_{KL} = \left(\begin{array}{r r r} 0 & -1 & 1\cr 0 & -1 & 1\cr -1 & 1 & 0\end{array}\right)\!,
\end{eqnarray}
where the last row in ${\bf F}_{CL}$ corresponds to the tree branch $C_{\rm sh}$ 
and the rightmost column in both ${\bf F}_{CL}$ and ${\bf F}_{KL}$ correspond 
to the loop closed by the chord $L_{\rm sh}$.
\begin{figure}[b]
\centerline{\includegraphics[width=8.5cm]{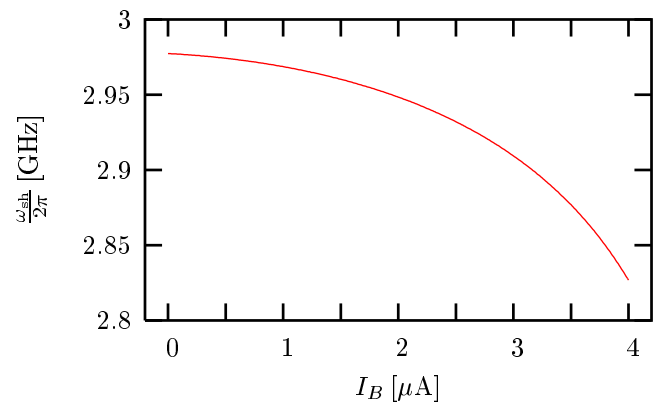}}
\caption{\label{plasma-fig}
Plasma frequency $\omega_{\rm sh}$ as a function of the applied bias current $I_B$.
The variation is due to the change the effective in Josephson inductances as
$I_B$ is varied.}
\end{figure}
Neglecting decoherence, the total Hamiltonian can be written as
\begin{equation}
  \label{totalH}
  {\cal H} = {\cal H}_S + {\cal H}_{\rm sh} + {\cal H}_{S,{\rm sh}},
\end{equation}
where ${\cal H}_S$, defined in Eq.~(\ref{Hamiltonian-S}), describes the
qubit and SQUID system.  The Hamiltonian of the plasmon mode can 
be brought into the second quantized form
\begin{equation}
  \label{plasmon1}
  {\cal H}_{\rm sh} = \frac{Q_{\rm sh}^2}{2C_{\rm sh}}
+\left(\frac{\Phi_0}{2\pi}\right)^2\frac{\varphi_{\rm sh}^2}{2L_{\rm t}}
= \hbar\omega_{\rm sh}\left(b^\dagger b +\frac{1}{2}\right),
\end{equation}
by introducing the resonance frequency $\omega_{\rm sh}=1/\sqrt{L_t C_{\rm sh}}$,
the total inductance (where the SQUID junctions have been linearized at the
operating point)
$L_t\simeq L_{\rm sh}+L'/4+L_J'/(\cos(\varphi_L)+\cos(\varphi_R))$,
and the creation and annihilation operators $b^\dagger$ and $b$, via 
\begin{equation}
\varphi_{\rm sh}= \frac{2\pi}{\Phi_0}\sqrt{\frac{\hbar}{2C_{\rm sh}\omega_{\rm sh}}}(b+b^\dagger) = 2\sqrt{\pi}\sqrt{\frac{Z_{\rm sh}}{R_Q}}(b+b^\dagger),
\label{b_operators}
\end{equation}
with the impedance $Z_{\rm sh}=\sqrt{L_t/C_{\rm sh}}$.
For the coupling between the qubit/SQUID system (the phases $\bphi$) and
the plasmon mode (the phase $\varphi_{\rm sh}$ associated with the 
charge on $C_{\rm sh}$, $Q_{\rm sh}=C_{\rm sh}\Phi_0\dot\varphi_{\rm sh}/2\pi$,
we obtain
\begin{equation}
  {\cal H}_{S,{\rm sh}} = \left(\frac{\Phi_0}{2\pi}\right)^2 \frac{1}{M_{\rm sh}}\varphi_{\rm sh}{\bf m}\cdot\bphi,
\label{Hlambda0}
\end{equation}
where ${\bf m}$ is given in Eq.~(\ref{m}) and
$M_{\rm sh} \approx L_{\rm sh}+L'/4$
(the exact expression for $M_{\rm sh}$ is a rational function of $L_{\rm sh}$ and the
circuit inductances which we will not display here).
Using Eq.~(\ref{b_operators}) and the semiclassical approximation
\begin{equation}
  \label{semiclappr}
  {\bf m}\cdot\bphi \approx -\frac{1}{2}\sigma_z {\bf m}\cdot\Delta\bphi + {\rm const.},
\end{equation}
we arrive at
\begin{equation}
  \label{plasmon3}
  {\cal H}_{S,{\rm sh}} = \lambda \sigma_z (b+b^\dagger),
\end{equation}
with the coupling strength
\begin{equation}
  \label{lambda}
  \lambda = -\sqrt{\pi}\left(\frac{\Phi_0}{2\pi}\right)^2 \sqrt{\frac{Z_{\rm sh}}{R_Q}}\frac{1}{M_{\rm sh}}{\bf m}\cdot\Delta\bphi.
\end{equation}
Note that this coupling vanishes along the decoupling line (Fig.~\ref{decoupling})
and also rapidly with the increase of $L_{\rm sh}$.

The complete two-level Hamiltonian then has the well-known Jaynes-Cummings form,
\begin{equation}
  {\cal H} =  \Delta \sigma_x + \epsilon \sigma_z
              + \hbar\omega_{\rm sh}\left(b^\dagger b +\frac{1}{2}\right) 
              + \lambda \sigma_z (b+b^\dagger). \label{Jaynes-Cummings}
\end{equation}
For the parameters in Ref.~\onlinecite{Delft-new}, $C_{\rm sh}=12\,{\rm pF}$
 and $L_{\rm sh}=170\,{\rm pH}$, we find
$\omega_{\rm sh}\approx2\pi\times 2.9 \,{\rm GHz}$ 
(see Fig.~\ref{plasma-fig}) and $Z_{\rm sh} = 5\,\Omega$, thus
$\sqrt{Z_{\rm sh}/R_Q}\approx 0.01$.
Note that the dependence of the Josephson inductance
(and thus of $L_t$ and $\omega_{\rm sh}$) on the state of the qubit
leads to an ac Stark shift term $\propto \sigma_z b^\dagger b$ which was
neglected in the coupling Hamiltonian Eq.~(\ref{Jaynes-Cummings}).

\begin{figure}
\vspace{5mm}
\centerline{\includegraphics[width=8.5cm]{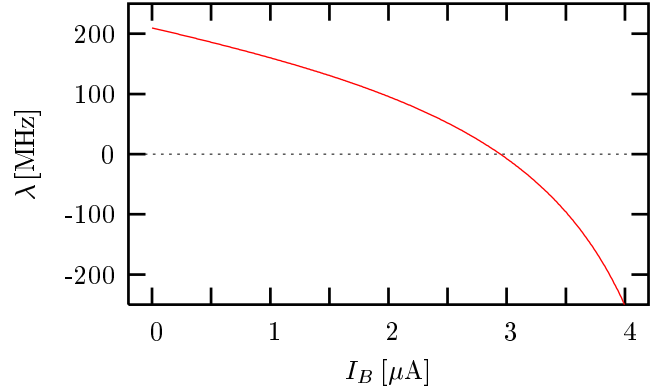}}
\caption{\label{lambda-fig}
Coupling constant $\lambda$ between the qubit and the plasmon mode.
The coupling disappears at the crossing with the decoupling line (Fig.~\ref{decoupling}),
i.e., when ${\bf m}\cdot\Delta\bphi=0$.}
\end{figure}

We find a coupling constant of $\lambda\approx 210\,{\rm MHz}$ at $I_B=0$.  
The coupling constant as a function of the bias
current $I_B$ is plotted in Fig.~\ref{lambda-fig}.
The relatively high values of $\lambda$ should allow the study
of the coupled dynamics of the qubit and the plasmon mode.
In particular, recently observed side resonances with the sum and difference
frequencies $E \pm \omega_{\rm sh}$ \cite{Chiorescu-coupling} 
can be explained in terms of the coupled dynamics, Eq.~(\ref{Jaynes-Cummings}).
Also, it should be possible to tune \emph{in-situ} the coupling to the plasmon mode
$\lambda$ at will, using pulsed bias currents.

\vspace{10mm}
\section{Discussion}
\label{Discussion}

We have found that the double-layer structure of SC circuits fabricated using
the shadow evaporation technique can drastically change the quantum dynamics
of the circuit due to the presence of unintended junctions.
In particular, the double-layer structure breaks the symmetry of the Delft 
qubit \cite{CNHM} (see Fig.~\ref{qubit-i}), and leads to relaxation and decoherence.  
We explain theoretically the observed compensation of the asymmetry at high $I_B$
\cite{Delft-new} 
and calculate the relaxation and decoherence times $T_1$ and $T_2$ of the qubit,
plotted in Fig.~\ref{times}. 
We find good quantitative agreement between theory and experiment in the 
value of the decoupling current $I_B$ where the relaxation and decoherence
times $T_1$ and $T_2$ reach their maximum.  In future qubit designs, the
asymmetry can be avoided by adding a fourth junction in series with the three
qubit junctions. It has already been demonstrated that this leads to a
shift of the maxima of $T_1$ and $T_2$ close to $I_B=0$, as theoretically
expected, and to an increase of the maximal values of $T_1$ and $T_2$
\cite{Delft-new}.

The asymmetry of the circuit also gives rise to an interesting coupling
between the qubit and an LC resonance in the external circuit (plasmon 
mode), which has been observed experimentally \cite{Chiorescu-coupling}, 
and which we have explained 
theoretically.  The coupling could potentially lead to interesting effects,
e.g., Rabi oscillations or entanglement between the qubit and the plasmon mode.

\section*{Acknowledgments}
GB and DPDV would like to acknowledge the hospitality of the Quantum Transport group
at TU Delft where this work was started.  DPDV was supported in part by the National
Security Agency and the Advanced Research and Development Activity through Army Research
Office contracts DAAD19-01-C-0056 and W911NF-04-C-0098.
PB acknowledges financial support
from a European Community Marie Curie fellowship.


\begin{thebibliography}{99}
\bibitem{fn1}
The semiclassical approximation accurately describes the double well
in the case where the states centered at the left and right minima
are well localized (see Ref.~\onlinecite{BKD}, Sec.~XIA).
\end{thebibliography}
\end{document}